\documentclass[aps,prl,showpacs,twocolumn,groupedaddress]{revtex4}  
\usepackage{graphicx}  
\usepackage{dcolumn}   
\usepackage{bm}        
\usepackage{amssymb}   

\begin{document}

\hspace{5.2in} \mbox{FERMILAB-Pub-04/215-E}

\title{A Search for the Flavor-Changing Neutral Current Decay $B^0_s\rightarrow \mu^{+}\mu^{-}$
in $p{\overline p}$ Collisions at $\sqrt s = 1.96$ TeV with the D\O \ Detector}
%
\author{                                                                      
V.M.~Abazov,$^{33}$                                                           
B.~Abbott,$^{70}$                                                             
M.~Abolins,$^{61}$                                                            
B.S.~Acharya,$^{27}$                                                          
M.~Adams,$^{48}$                                                              
T.~Adams,$^{46}$                                                              
M.~Agelou,$^{17}$                                                             
J.-L.~Agram,$^{18}$                                                           
S.H.~Ahn,$^{29}$                                                              
M.~Ahsan,$^{55}$
% ...
\\                                                                            
\vskip 0.30cm                                                                 
\centerline{(D\O\ Collaboration)}                                             
\vskip 0.30cm                                                                 
}                                                                             
\address{                                                                     
\centerline{$^{1}$Universidad de Buenos Aires, Buenos Aires, Argentina}       
\centerline{$^{2}$LAFEX, Centro Brasileiro de Pesquisas F{\'\i}sicas,         
                  Rio de Janeiro, Brazil}                                     
\centerline{$^{3}$Universidade do Estado do Rio de Janeiro,                   
                  Rio de Janeiro, Brazil}                                     
\centerline{$^{4}$Instituto de F\'{\i}sica Te\'orica, Universidade            
                  Estadual Paulista, S\~ao Paulo, Brazil}                     
\centerline{$^{5}$Simon Fraser University, Burnaby, Canada, University of     
                  Alberta, Edmonton, Canada,}                                 
\centerline{McGill University, Montreal, Canada and York University,          
                  Toronto, Canada}                                            
\centerline{$^{6}$Institute of High Energy Physics, Beijing,                  
                  People's Republic of China}                                 
\centerline{$^{7}$Universidad de los Andes, Bogot\'{a}, Colombia}             
\centerline{$^{8}$Charles University, Center for Particle Physics,            
                  Prague, Czech Republic}                                     
\centerline{$^{9}$Czech Technical University, Prague, Czech Republic}         
\centerline{$^{10}$Institute of Physics, Academy of Sciences, Center          
                  for Particle Physics, Prague, Czech Republic}               
\centerline{$^{11}$Universidad San Francisco de Quito, Quito, Ecuador}        
\centerline{$^{12}$Laboratoire de Physique Corpusculaire, IN2P3-CNRS,         
                 Universit\'e Blaise Pascal, Clermont-Ferrand, France}        
\centerline{$^{13}$Laboratoire de Physique Subatomique et de Cosmologie,      
                  IN2P3-CNRS, Universite de Grenoble 1, Grenoble, France}     
\centerline{$^{14}$CPPM, IN2P3-CNRS, Universit\'e de la M\'editerran\'ee,     
                  Marseille, France}                                          
\centerline{$^{15}$Laboratoire de l'Acc\'el\'erateur Lin\'eaire,              
                  IN2P3-CNRS, Orsay, France}                                  
\centerline{$^{16}$LPNHE, Universit\'es Paris VI and VII, IN2P3-CNRS,         
                  Paris, France}                                              
\centerline{$^{17}$DAPNIA/Service de Physique des Particules, CEA, Saclay,    
                  France}                                                     
\centerline{$^{18}$IReS, IN2P3-CNRS, Universit\'e Louis Pasteur, Strasbourg,  
                  France and Universit\'e de Haute Alsace, Mulhouse, France}  
\centerline{$^{19}$Institut de Physique Nucl\'eaire de Lyon, IN2P3-CNRS,      
                   Universit\'e Claude Bernard, Villeurbanne, France}         
\centerline{$^{20}$RWTH Aachen, III. Physikalisches Institut A,               
                   Aachen, Germany}                                           
\centerline{$^{21}$Universit{\"a}t Bonn, Physikalisches Institut,             
                  Bonn, Germany}                                              
\centerline{$^{22}$Universit{\"a}t Freiburg, Physikalisches Institut,         
                  Freiburg, Germany}                                          
\centerline{$^{23}$Universit{\"a}t Mainz, Institut f{\"u}r Physik,            
                  Mainz, Germany}                                             
\centerline{$^{24}$Ludwig-Maximilians-Universit{\"a}t M{\"u}nchen,            
                   M{\"u}nchen, Germany}                                      
\centerline{$^{25}$Fachbereich Physik, University of Wuppertal,               
                   Wuppertal, Germany}                                        
\centerline{$^{26}$Panjab University, Chandigarh, India}                      
\centerline{$^{27}$Tata Institute of Fundamental Research, Mumbai, India}     
\centerline{$^{28}$University College Dublin, Dublin, Ireland}                
\centerline{$^{29}$Korea Detector Laboratory, Korea University,               
                   Seoul, Korea}                                              
\centerline{$^{30}$CINVESTAV, Mexico City, Mexico}                            
\centerline{$^{31}$FOM-Institute NIKHEF and University of                     
                  Amsterdam/NIKHEF, Amsterdam, The Netherlands}               
\centerline{$^{32}$University of Nijmegen/NIKHEF, Nijmegen, The               
                  Netherlands}                                                
\centerline{$^{33}$Joint Institute for Nuclear Research, Dubna, Russia}       
\centerline{$^{34}$Institute for Theoretical and Experimental Physics,        
                  Moscow, Russia}                                             
\centerline{$^{35}$Moscow State University, Moscow, Russia}                   
\centerline{$^{36}$Institute for High Energy Physics, Protvino, Russia}       
\centerline{$^{37}$Petersburg Nuclear Physics Institute,                      
                   St. Petersburg, Russia}                                    
\centerline{$^{38}$Lund University, Lund, Sweden, Royal Institute of          
                   Technology and Stockholm University, Stockholm,            
                   Sweden and}                                                
\centerline{Uppsala University, Uppsala, Sweden}                              
\centerline{$^{39}$Lancaster University, Lancaster, United Kingdom}           
\centerline{$^{40}$Imperial College, London, United Kingdom}                  
\centerline{$^{41}$University of Manchester, Manchester, United Kingdom}      
\centerline{$^{42}$University of Arizona, Tucson, Arizona 85721, USA}         
\centerline{$^{43}$Lawrence Berkeley National Laboratory and University of    
                  California, Berkeley, California 94720, USA}                
\centerline{$^{44}$California State University, Fresno, California 93740, USA}
\centerline{$^{45}$University of California, Riverside, California 92521, USA}
\centerline{$^{46}$Florida State University, Tallahassee, Florida 32306, USA} 
\centerline{$^{47}$Fermi National Accelerator Laboratory, Batavia,            
                   Illinois 60510, USA}                                       
\centerline{$^{48}$University of Illinois at Chicago, Chicago,                
                   Illinois 60607, USA}                                       
\centerline{$^{49}$Northern Illinois University, DeKalb, Illinois 60115, USA} 
\centerline{$^{50}$Northwestern University, Evanston, Illinois 60208, USA}    
\centerline{$^{51}$Indiana University, Bloomington, Indiana 47405, USA}       
\centerline{$^{52}$University of Notre Dame, Notre Dame, Indiana 46556, USA}  
\centerline{$^{53}$Iowa State University, Ames, Iowa 50011, USA}              
\centerline{$^{54}$University of Kansas, Lawrence, Kansas 66045, USA}         
\centerline{$^{55}$Kansas State University, Manhattan, Kansas 66506, USA}     
\centerline{$^{56}$Louisiana Tech University, Ruston, Louisiana 71272, USA}   
\centerline{$^{57}$University of Maryland, College Park, Maryland 20742, USA} 
\centerline{$^{58}$Boston University, Boston, Massachusetts 02215, USA}       
\centerline{$^{59}$Northeastern University, Boston, Massachusetts 02115, USA} 
\centerline{$^{60}$University of Michigan, Ann Arbor, Michigan 48109, USA}    
\centerline{$^{61}$Michigan State University, East Lansing, Michigan 48824,   
                   USA}                                                       
\centerline{$^{62}$University of Mississippi, University, Mississippi 38677,  
                   USA}                                                       
\centerline{$^{63}$University of Nebraska, Lincoln, Nebraska 68588, USA}      
\centerline{$^{64}$Princeton University, Princeton, New Jersey 08544, USA}    
\centerline{$^{65}$Columbia University, New York, New York 10027, USA}        
\centerline{$^{66}$University of Rochester, Rochester, New York 14627, USA}   
\centerline{$^{67}$State University of New York, Stony Brook,                 
                   New York 11794, USA}                                       
\centerline{$^{68}$Brookhaven National Laboratory, Upton, New York 11973, USA}
\centerline{$^{69}$Langston University, Langston, Oklahoma 73050, USA}        
\centerline{$^{70}$University of Oklahoma, Norman, Oklahoma 73019, USA}       
\centerline{$^{71}$Brown University, Providence, Rhode Island 02912, USA}     
\centerline{$^{72}$University of Texas, Arlington, Texas 76019, USA}          
\centerline{$^{73}$Southern Methodist University, Dallas, Texas 75275, USA}   
\centerline{$^{74}$Rice University, Houston, Texas 77005, USA}                
\centerline{$^{75}$University of Virginia, Charlottesville, Virginia 22901,   
                   USA}                                                       
\centerline{$^{76}$University of Washington, Seattle, Washington 98195, USA}  
}                                                                             
\date{\today}

\begin{abstract}

We present the results of a search for the flavor-changing neutral current
decay $B^0_s \rightarrow \mu^+ \mu^-$ using a data set with integrated luminosity of 240~pb$^{-1}$ of $p\bar{p}$ collisions
at $\sqrt{s}=1.96$~TeV collected with the D\O~detector in Run II of the
Fermilab Tevatron collider. We find the upper limit on the branching fraction to
be ${\cal B}(B^0_s \rightarrow \mu^+ \mu^-) \leq 5.0\times
10^{-7}$ at the 95\% C.L. assuming no contributions from the decay $B^0_d\rightarrow \mu^+\mu^-$
in the signal region. This limit is the most stringent upper bound on the branching fraction $B^0_s\rightarrow \mu^+\mu^-$ to date.
\end{abstract}

\maketitle

The purely leptonic decays $B_{d,s}^0 \rightarrow \mu^+
\mu^-$~\cite{conjugated} are flavor-changing neutral current (FCNC) processes.
In the standard model (SM), these decays are forbidden at the tree level and
proceed at a very low rate through higher-order diagrams.
The SM leptonic branching fractions (${\cal B}$) were calculated
including QCD corrections in Ref.~\cite{sm_ref2}. The latest SM prediction~\cite{sm_ref3}
is ${\cal B}(B^0_s \rightarrow \mu^+ \mu^-)=(3.42\pm 0.54)\times
10^{-9}$, where the error is dominated by non-perturbative uncertainties. The
leptonic branching fraction of the $B_d^0$ decay is suppressed by CKM matrix elements $|V_{td}/V_{ts}|^2$
leading to a predicted SM branching fraction of $(1.00\pm0.14)\times 10^{-10}$.
The best existing experimental bound for the branching fraction
of $B^0_s$ $(B^0_d)$ is presently
${\cal B}(B^0_s \, (B^0_d) \rightarrow \mu^+\mu^-)<7.5\times 10^{-7}\, (1.9\times 10^{-7})$ at the 95\% C.L.~\cite{cdfII}.

The decay amplitude of $B^0_{d,s} \rightarrow \mu^+ \mu^-$ can be
significantly enhanced in some extensions of the SM. For instance, in
the type-II two-Higgs-doublet model (2HDM) the branching fraction depends
only on the charged Higgs mass $M_{H^+}$ and $\tan \beta$, the ratio of the two neutral Higgs
field vacuum expectation values, with the branching fraction growing like $(\tan \beta)^4$~\cite{nierste}. In the
minimal supersymmetric standard model (MSSM), however, ${\cal B}(B^0_s \rightarrow \mu^+ \mu^-) \propto(\tan
\beta)^6$, leading to an enhancement of up to three orders of
magnitude~\cite{dedes} compared to the SM, even if the MSSM with minimal
flavor violation (MFV) is considered, i.e., the CKM matrix is the only
source of flavor violation. An observation of $B^0_{s} \rightarrow \mu^+ \mu^-$ would then immediately lead to an upper bound on the heaviest mass in the MSSM Higgs sector~\cite{dedes_huffman} if MFV applies.
In minimal supergravity models, an enhancement of ${\cal
B}(B^0_s \rightarrow \mu^+ \mu^-)$ is correlated~\cite{nierste_prl}
with a sizeable positive shift in $(g-2)_\mu$ that also requires
large $\tan \beta$. A large value of $\tan\beta$ is theoretically well-motivated by
grand unified theories (GUT) based on minimal SO(10). These models predict
large enhancements of ${\cal B}(B^0_s \rightarrow \mu^+ \mu^-)$ as
well~\cite{nierste_prl,s10}. Finally, FCNC decays of $B^0_{d,s}$ are also sensitive
to supersymmetric models with non-minimal flavor violation structures such as the generic
MSSM~\cite{bobeth} and $R$-parity violating supersymmetry~\cite{rp_susy}.

In this Letter we report on a search for the decay $B^0_s\rightarrow \mu^{+}\mu^{-}$ using a data set of integrated luminosity of 240~pb$^{-1}$ recorded with the D\O \ detector in the years 2002--2004. Our mass resolution is not sufficient to readily separate $B^0_s$ from $B^0_d$ leptonic decays. For the final calculation of the upper limit on ${\cal B}(B^0_s\rightarrow \mu^+\mu^-)$ we assumed that there is no contribution from $B^0_d\rightarrow
\mu^{+}\mu^{-}$ decays in our search region due to its
suppression by $|V_{td}/V_{ts}|^2$, which holds in all models with
MFV.

The D\O\ detector is described in detail elsewhere~\cite{run2det}. The main
elements, relevant for this analysis, are the central tracking and muon detector systems.
The central tracking system consists of a silicon microstrip tracker (SMT) and a central
fiber tracker (CFT), both located within a 2~T superconducting
solenoidal magnet. Located outside the calorimeter, the muon detector consists of a layer of
tracking detectors and
scintillation trigger counters in front of toroidal magnets (1.8~T),
followed by two more similar layers behind the toroids, allowing for
efficient muon detection out to $\eta$ of about $\pm 2$, where $\eta=-\ln[\tan(\theta/2)]$ is the pseudorapidity and $\theta$ is the polar angle measured relative to the proton beam direction.

Four versions of dimuon triggers were used in the data selection of
this analysis.  A trigger simulation was used to estimate the trigger
efficiency for the signal and normalization samples.
These efficiencies were also checked with data
samples collected with single muon triggers. Event preselection
started by requiring two muons of opposite charge to be identified by extrapolating
charged tracks reconstructed in the
central tracking detectors to the muon detectors, and matching them
with hits in the latter. The muons had to form a common secondary 3D-vertex with
an invariant mass $m(\mu^+\mu^-)$ between 4.5 and 7.0~GeV/$c^2$ and a $\chi^2$ per degree of freedom of $\chi^2/\mbox{d.o.f}<10$.
Each muon was required to have $p_T>2.5$~GeV/$c$ and $|\eta|<2.0$. Tracks that
were matched to each muon were required to have at least three hits in
the SMT and at least four hits in the CFT. To select well-measured secondary
vertices, we determined the two-dimensional decay length $L_{xy}$ in the plane
transverse to the beamline, and required the uncertainty $\delta
L_{xy}$ to be less than 0.15~mm. $L_{xy}$ was calculated
as $L_{xy}=\frac{\vec{l}_{vtx}\cdot \vec{p}_T^B}{p_T^B}$, where $p_T^B$ is the
transverse momentum of the candidate $B^0_s$ and $\vec{l}_{vtx}$ represents
the vector pointing from the primary
vertex to the secondary vertex. The error on the transverse decay
length, $\delta L_{xy}$, was calculated by taking into account the
uncertainties in both the primary and secondary vertex positions. The
primary vertex itself was found for each event using a beam-spot
constrained fit as described in Ref.~\cite{delphi}. To ensure a similar
$p_T$ dependence of the $\mu^+\mu^-$ system in the signal and in
the normalization channel, $p_T^B$ had to be greater than 5~GeV/$c$. A total
of 38,167 events survive these preselection requirements.
The effects of these criteria on the number of events are shown
in Table~\ref{tab_eff}.

\begin{table}[htbp]
\caption{Number of candidate events in data satisfying successive preselection requirements.}
\small
\begin{center}
\begin{tabular}{lcc}\hline\hline

Variable &  Requirement  & \# Candidates   \\
\hline
   Mass (GeV/$c^2$) & 4.5 $<m_{\mu^+ \mu^-} <$7.0 & 405,307\\
   Muon quality &  & 234,792 \\
   $\chi^2/{\rm d.o.f}$ of vertex  & $< 10$      & 146,982 \\
   Muon $p_T$ (GeV/$c$)      &  $>$ 2.5  & 129,558 \\
   Muon $|\eta|$                 & $<$ 2.0  & 125,679 \\
   Tracking hits    &   CFT $\ge$ 4, SMT $\ge$ 3     & 92,678 \\
      $\delta L_{xy}$ (mm)&    $<$  0.15      & 90,935   \\
   $B^0_s$ candidate $p_T^B$ (GeV/$c$) &       $>$ 5.0       & 38,167 \\ \hline\hline
\end{tabular}
\label{tab_eff}
\end{center}
\end{table}
For the final event selection, we required the candidate events to pass
additional criteria. The long lifetime of the $B^0_s$ mesons allows us
to reject random combinatoric background. We therefore used the decay length significance
$L_{xy}/\delta L_{xy}$ as one of the discriminating variables, since it gives better
discriminating power than the transverse decay length alone, as large values of $L_{xy}$ may originate
due to large uncertainties.

The fragmentation characteristics of the $b$ quark are such that most of its momentum is carried by the $b$ hadron. Thus the number of extra tracks near the $B^0_s$ candidate tends to be small. The second
discriminant was therefore an isolation variable, ${\cal I}$, of the muon pair, defined as:
\begin{equation}
   {\cal I}  = \frac{|\vec{p}(\mu^+\mu^-)|}{|\vec{p}(\mu^+\mu^-)|+ \sum\limits_{{\rm track}\,i \neq B}{ p_i(\Delta {\cal R} < 1)} }.\nonumber
\end{equation}
Here, $\sum\limits_{{\rm track}\,i \neq B}{ p_i}$ is the scalar sum over
all tracks excluding the muon pair within a cone of $\Delta {\cal R} < 1$
around the momentum vector $\vec{p}(\mu^+ \mu^-)$ of the muon pair where
$\Delta {\cal R} = \sqrt{(\Delta\phi)^2 + (\Delta\eta)^2 }$.


The final discriminating variable was the pointing angle $\alpha$,
defined as the angle between the momentum vector $\vec{p}(\mu^+
\mu^-)$ of the muon pair and the vector $\vec{l}_{vtx}$ between
the primary and secondary vertices. This requirement ensured consistency between the
direction of the decay vertex and the momentum vector of the $B^0_s$
candidate.

An optimization based on these discriminating variables was done on
signal Monte Carlo (MC) events in the $B_s^0$ mass region $4.53 < M_{\mu^+\mu^-} <
6.15$~GeV/$c^2$ with $m_{B^0_s}=5369.6\pm 2.4$~MeV/$c^2$~\cite{pdg}
and on data events in regions outside the signal window, i.e., in the sidebands. The mass scale
throughout this analysis is shifted downward with respect to the world average $B^0_s$
mass by 30~MeV/$c^2$ to compensate for the shift in the momentum
scale of the D\O \ tracking system. The mass shift was found by linear interpolation
to the $B^0_s$ mass of the measured mass shifts between the $J/\psi$ and the $\Upsilon$
resonances relative to their world average values~\cite{pdg}. The mass shift is smaller than the
MC predicted mass resolution for two-body decays of $\sigma=90$~MeV/$c^2$ at the $B^0_s$ mass.

In order to avoid biasing the optimization procedure, data candidates in the signal mass region were not
examined until completion of the analysis, and events in the sideband regions around
the $B^0_s$ mass were used instead. The start (end) of the upper (lower) sideband was
chosen such that they were at least $3\sigma$ (270 MeV/$c^2$) away from
the $B^0_s$ mass. The widths of the sidebands used
for background estimation were chosen to be $6\sigma$ each. The size of the blind signal region was $\pm 3\sigma$ around the $B^0_s$ mass. To determine the limit on the branching fraction, we used a smaller mass region of $\pm 2\sigma$.

\begin{figure*}
\begin{minipage}[l]{0.31\textwidth}
  \includegraphics*[width=\linewidth]{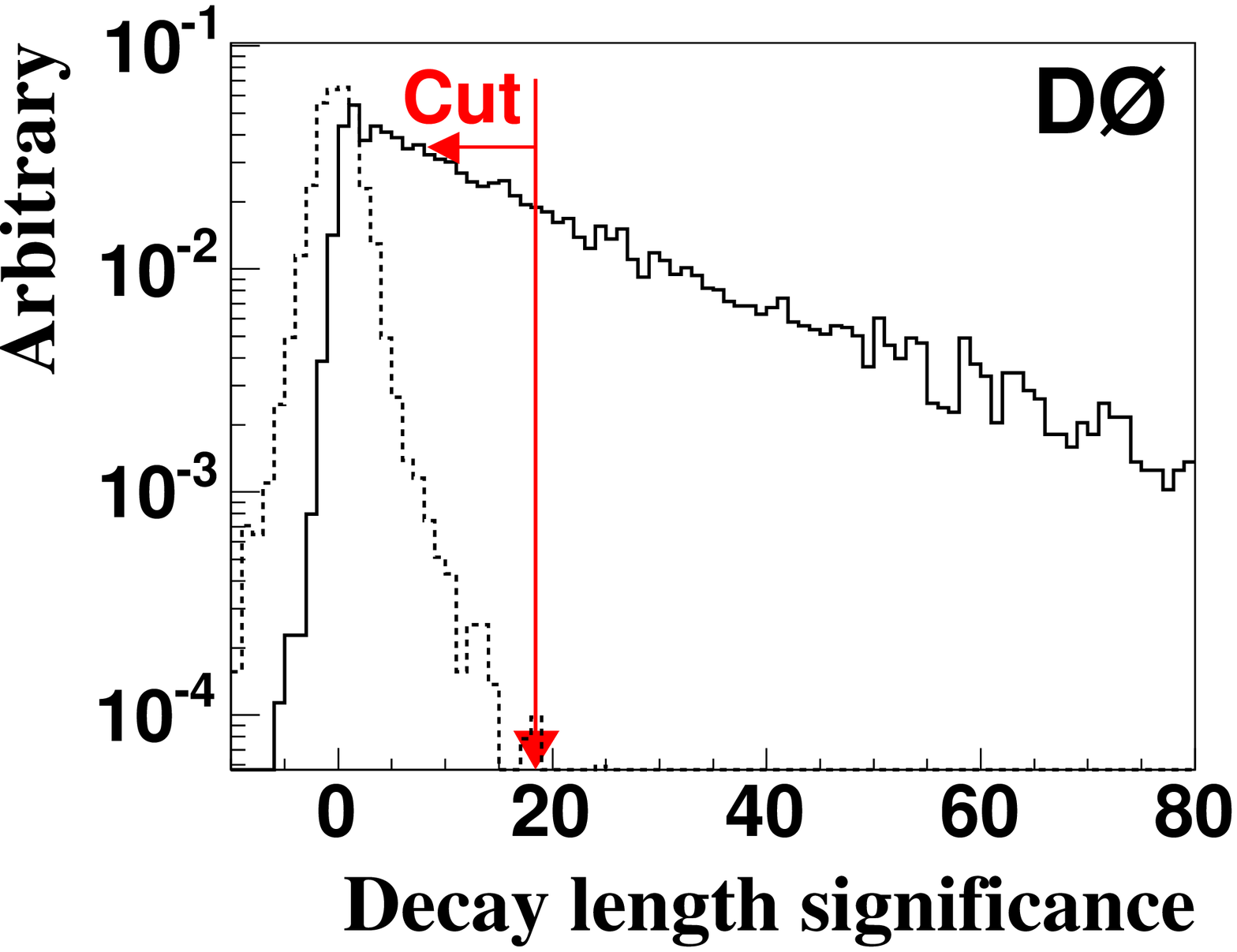}
\end{minipage}
\begin{minipage}[l]{0.31\textwidth}
  \includegraphics*[width=\linewidth]{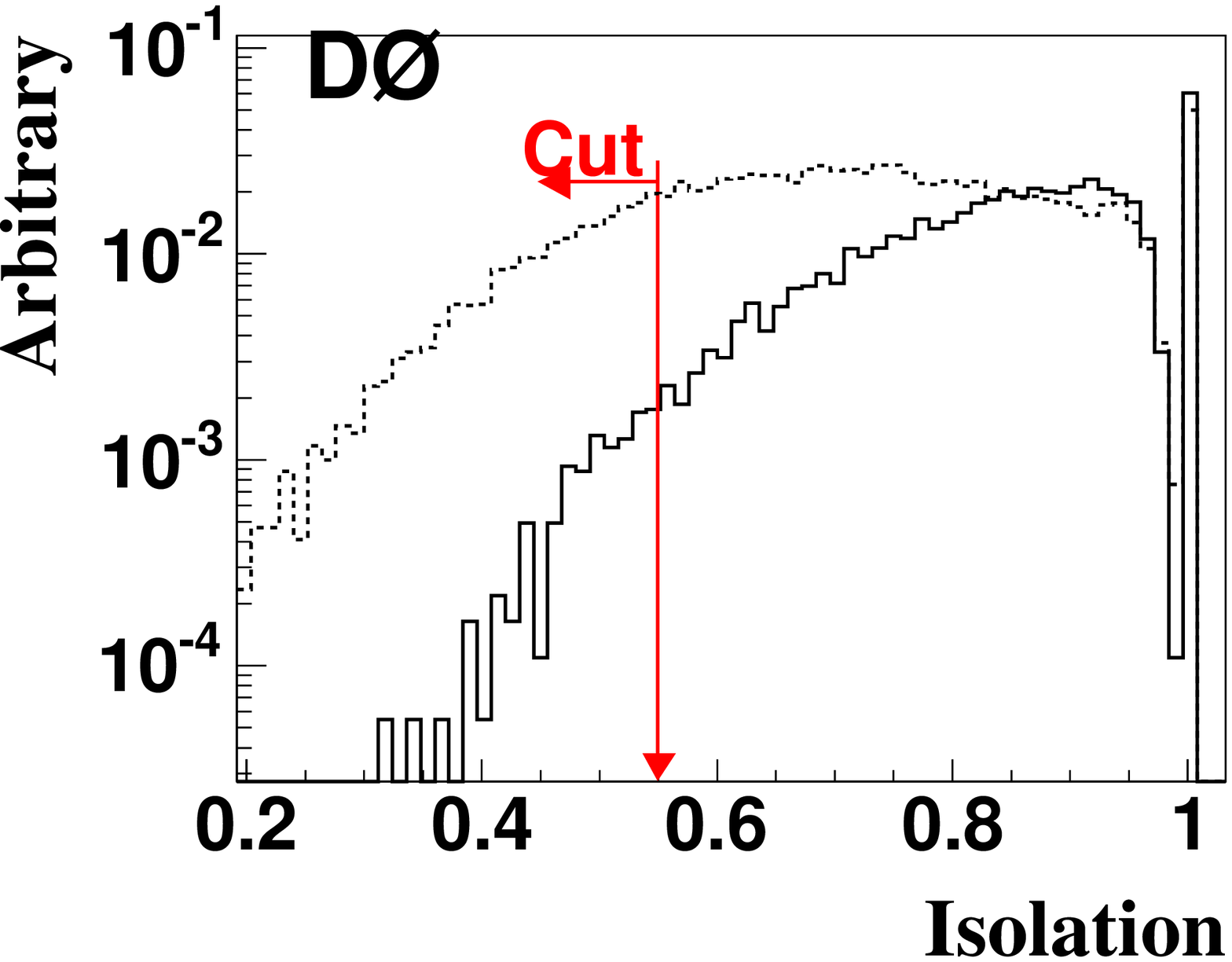}
\end{minipage}
\begin{minipage}[l]{0.31\textwidth}
  \includegraphics*[width=\linewidth]{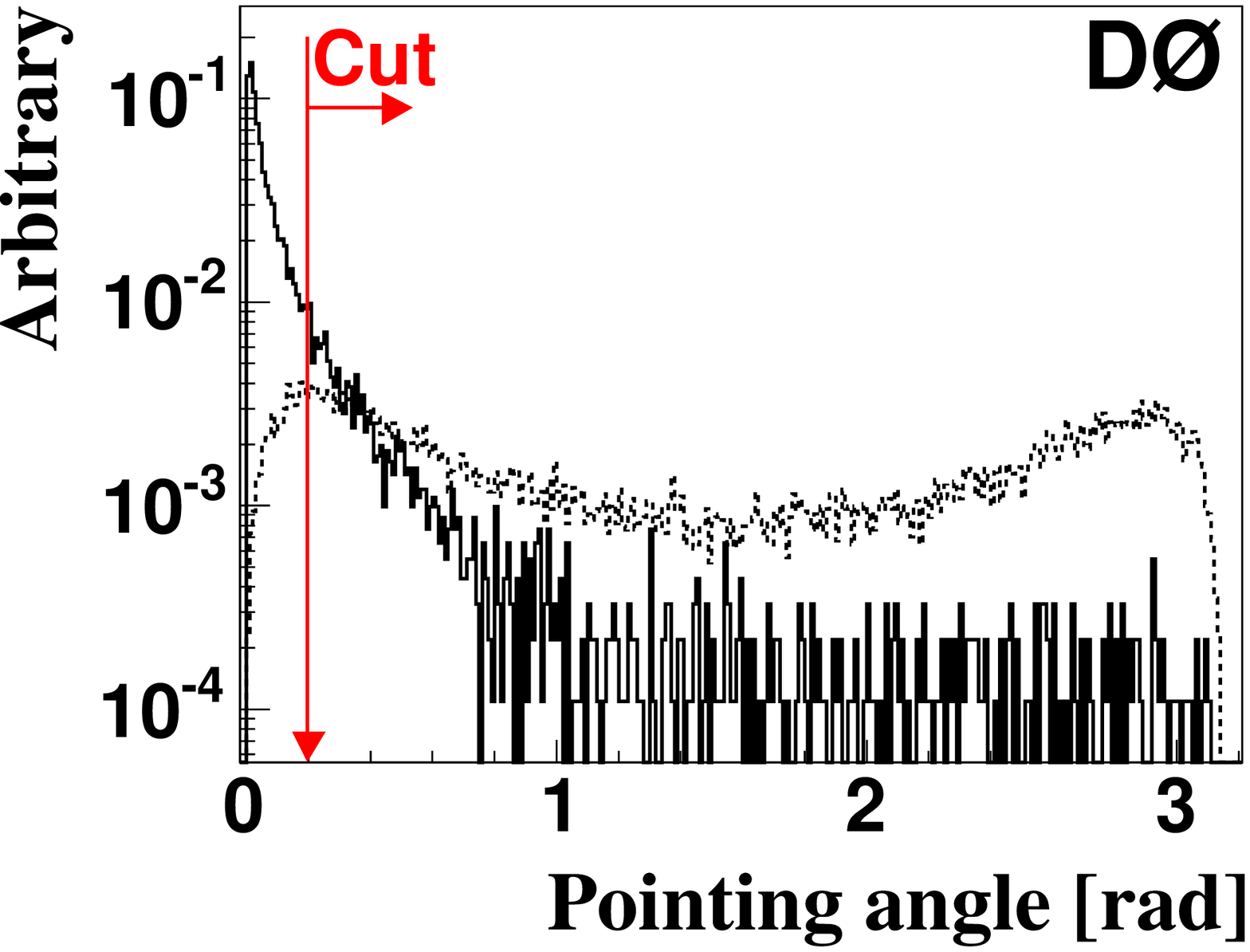}
\end{minipage}
  \caption{\label{fig_disc}Discriminating variables after the preselection for signal MC (solid line) and data events (dashed line)
  from the sidebands. The arrows indicate the discriminating values that were obtained after optimization. The normalization is done on the number of signal MC and sideband data events after preselection.}
\end{figure*}

A random-grid search~\cite{rgs} and an optimization procedure~\cite{punzi} were
used to find the optimal values of
the discriminating variables, by maximizing the variable
$P=\epsilon_{\mu\mu}^{B^0_s}/(a/2+\sqrt{N_{\rm back}})$.
Here, $\epsilon_{\mu\mu}^{B^0_s}$ is the reconstruction efficiency of the
signal events relative to the preselection (estimated using MC), and
$N_{\rm back}$ is the expected number of background events interpolated
from the sidebands. The constant $a$ is the number of standard deviations
corresponding to the confidence level at which the signal hypothesis
is tested. This constant $a$ was set to 2.0, corresponding to about the 95\% C.L.
Figure~\ref{fig_disc} shows the distribution of the three discriminating variables after the
preselection for signal MC events and data in the sideband regions. After optimization, we found the following
values for the discriminating variables and MC signal efficiencies relative to the preselected sample: $L_{xy}/\delta L_{xy}>18.5$ (47.5\%), ${\cal I}>0.56$ (97.4\%), and $\alpha<0.2$~rad (83.4\%). The combined efficiency for signal events to survive these three additional selection criteria, as measured relative to preselection criteria,
is (38.6$\pm$0.7)\%, where the error is due to limited MC statistics.
A linear extrapolation of the sideband population for the whole data
sample into the ($\pm$180~MeV/$c^2$) signal region yields an expected
number of 3.7$\pm$1.1 background events.

\begin{figure}
  \includegraphics*[width=\linewidth]{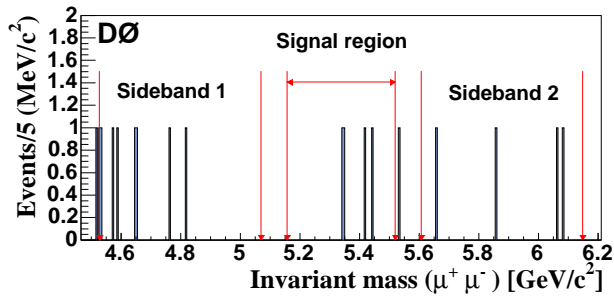}
  \caption{\label{fig_background_standard}Invariant mass of the remaining events of the full data sample after optimized requirements on the discriminating variables.}
\end{figure}

Upon examining the data in all mass regions, four events
are observed in the  signal region, entirely
consistent with the background events as
estimated from sidebands. We examined the four observed events
in detail by studying various kinematic variables, e.g., $p_T$
of the muons, isolation, etc., and found them to be compatible with background events.
Figure~\ref{fig_background_standard} shows the
remaining events populating the lower and upper sidebands
as well as the signal region almost equally.

In the absence of an apparent signal, a limit on the
branching fraction ${\cal B}(B^0_s\rightarrow \mu^+\mu^-)$ can be computed by normalizing the upper limit on
the number of events in the $B^0_s$ signal region to the number of reconstructed $B^{\pm}\rightarrow J/\psi\,K^{\pm}$ events:

\begin{equation}
\label{eq_limit2}
{\cal B}(B^0_s\rightarrow \mu^+\mu^-) \leq
\frac{N_{\rm ul}}{N_{B^\pm}}\cdot\frac{\epsilon_{\mu\mu K}^{B^\pm}}{\epsilon_{\mu\mu}^{B^0_s}}\cdot
\frac{{\cal B}(B^{\pm}\rightarrow J/\psi(\mu^+\mu^-) K^{\pm})}{\frac{f_{b\rightarrow B_{s}}}{f_{b \rightarrow B_{u,d}}}+ R\cdot
\frac{\epsilon_{\mu\mu}^{B^0_d}}{\epsilon_{\mu\mu}^{B^0_s}}},
\end{equation}
where

\begin{itemize}
\item $N_{\rm ul}$ is the upper limit on the number of signal decays estimated from the number of observed events and expected background events.
\item $N_{B^{\pm}}$ is the number of observed $B^{\pm}\rightarrow J/\psi\,K^{\pm}$ events.
\item $\epsilon_{\mu\mu}^{B^0_s}$ and $\epsilon_{\mu\mu K}^{B^\pm}$ are the
  efficiencies of the signal and normalization channels, obtained from
  MC simulations.
\item ${\cal B}(B^{\pm}\rightarrow J/\psi(\mu^+\mu^-) K^{\pm})$ is the product of the branching fractions ${\cal B}(B^\pm \rightarrow J/\psi\,K^\pm)=(1.00\pm 0.04)\times 10
  ^{-3}$ and ${\cal B}( J/\psi \rightarrow \mu^+ \mu^-)=(5.88\pm 0.10)\times 10^{-2}$~\cite{pdg}.
\item $f_{b\rightarrow B^0_{s}}/f_{b \rightarrow B_{u,d}}=0.270\pm 0.034$ is
  the fragmentation ratio of a $\bar{b}$ quark producing a $B^0_s$ and a $B_{u,d}$ meson. This ratio has been calculated using the latest world average fragmentation
values~\cite{pdg} for $B^0_s$ and $B_{u,d}$ mesons, where the uncertainty on the ratio is conservatively calculated assuming a full anti-correlation among the individual $B_{u,d}$ and $B^0_s$ fragmentation uncertainties.
\item $R\cdot \epsilon_{\mu\mu}^{B^0_d}/\epsilon_{\mu\mu}^{B^0_s}$ is the branching fraction ratio $R={\cal B}(B^0_d)/{\cal B}(B^0_s)$ of $B^0_{d,s}$ mesons decaying into
two muons multiplied by the total detection efficiency ratio~\cite{eff_ratio}. Any non-negligible contribution due to $B^0_d$ decays ($R>0$) would
make the limit on the branching fraction ${\cal
B}(B^0_s\rightarrow \mu^+\mu^-)$ as given in Eq.~(\ref{eq_limit2})
smaller. Our limit presented for ${\cal B}(B^0_s\rightarrow
\mu^+\mu^-)$ is therefore conservative.

\end{itemize}

Using the $B^{\pm}\rightarrow J/\psi\,K^{\pm}$ mode~\cite{norm} has the advantage that the efficiencies to detect the $\mu^+\mu^-$ system in signal and normalization
events are similar, and systematic effects tend to cancel.
A pure sample of $B^{\pm}\rightarrow J/\psi\,K^{\pm}$ events was obtained
by applying the following selection criteria. The mass-constrained vertex fit of
the two muons to form a $J/\psi$ was required to have a
$\chi^2/\mbox{d.o.f.}<10$, similar to the $\mu^+\mu^-$ vertex
criterion in the $B^0_s\rightarrow \mu^+\mu^-$ search. The combined vertex fit of the $J/\psi$
and the additional $K^{\pm}$ ($p_T(K^{\pm})>0.9$~GeV/$c$) had to have $\chi^2<20$ for three {\rm
d.o.f.}. The requirements on the three discriminating variables were also applied.
\begin{figure}
  \includegraphics*[width=\linewidth]{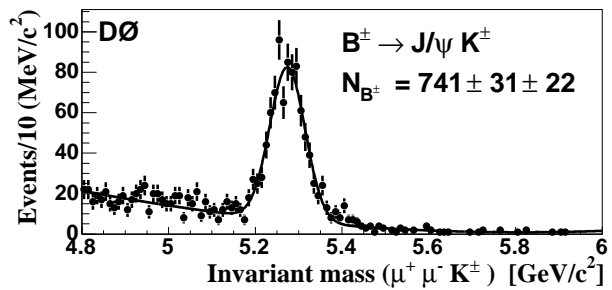}
  \caption{\label{fig_norm}Invariant mass distribution for candidates in the normalization channel $B^{\pm}\rightarrow J/\psi K^{\pm}$.}
\end{figure}
The mass spectrum of the reconstructed $B^{\pm}\rightarrow J/\psi\,
K^{\pm}$ for the full data sample after all analysis requirements is shown in
Fig.~\ref{fig_norm}. A fit using a Gaussian for the signal
and a second order polynomial for the background yields $741\,\pm \,31\,({\rm stat}) \pm
\,22 \,({\rm sys})$ $B^{\pm}$ candidates, where the systematic uncertainty was estimated by varying the
fit range, background and signal shape hypotheses.

The $p_T$ distribution of the $B^\pm$ in data has a slightly harder
spectrum than that from MC. Therefore, MC events of the signal and normalization channels
have been reweighted accordingly.
In addition, the observed widths of known $\mu^+ \mu^-$ resonances ($J/\psi$ and $\Upsilon(1S)$)
are (27$\pm$4)\% larger than predicted by MC.
The $\pm 2 \sigma$ signal mass region using the MC mass resolution
therefore corresponds to $\pm 1.58 \sigma$ when the data mass
resolution is considered, and the efficiency is corrected accordingly.
To within errors, the MC correctly reproduces the efficiency of the
cuts on the discriminating variables when applied to the normalization channel.

The final corrected value for the efficiency ratio
is then given by $\epsilon_{\mu\mu K}^{B^\pm}/\epsilon_{\mu\mu}^{B^0_s}=0.247\,\pm \,0.009\,({\rm stat}) \pm \,0.017\, ({\rm sys})$, where the first
uncertainty is due to limited MC statistics and the second accounts
for the $B^{\pm}/B^0_s$ lifetime ratio uncertainties and
for uncertainties in data/MC differences. These differences include the $p_T$-dependent reweighting of MC events, signal mass width, the kaon track reconstruction efficiency and the effects of different trigger and muon identification efficiencies.
All systematic uncertainties entering the
calculation of the branching fraction limit are listed in
Table~\ref{sys}.
\begin{table}
\caption{Relative uncertainties used in the calculation of an upper limit of ${\cal B}(B^0_s\rightarrow \mu^+\mu^-)$.}
\begin{center}
\begin{tabular}{lc}\hline\hline\label{sys}
   Source & Relative Uncertainty $[\%]$ \\
\hline
$\epsilon_{\mu\mu K}^{B^\pm}/\epsilon_{\mu\mu}^{B^0_s}$ & 7.7 \\
Number of $B^\pm \rightarrow J/\psi K^\pm$ events & 5.1 \\
${\cal B}(B^\pm \rightarrow J/\psi K^\pm)$ & 4.0\\
${\cal B}( J/\psi \rightarrow \mu^+ \mu^- )$ & 1.7\\
$f_{b \rightarrow B^0_s}/f_{b \rightarrow B^{0}_{u,d}}$  & 12.7\\
Background uncertainty & 29.7 \\
   \hline\hline
\end{tabular}
\end{center}
\end{table}

The statistical uncertainties on the background expectation, as well as
the uncertainties on the efficiencies can be included
into the limit calculation by integrating over
probability functions that parameterize the uncertainties. We have
used a prescription~\cite{conrad} to construct a confidence interval with the
Feldman and Cousins ordering scheme.
The expected background was modeled as a Gaussian distribution with its mean value equal to the expected number
of background events and its standard deviation equal to the background
uncertainty. The uncertainty on the number of $B^{\pm}$ events as well as the uncertainties on
the fragmentation ratio and branching fractions for $B^{\pm}\rightarrow J/\psi(\mu^+\mu^-)\, K^{\pm}$ were added in quadrature to the efficiency uncertainties and parameterized as a Gaussian distribution. The resulting branching fraction limit~\cite{limit} including all the statistical and systematic uncertainties at a 95\% (90\%) C.L. is given by
\begin{center}
${\cal B}(B^0_s \rightarrow \mu^+ \mu^-) \leq 5.0\, \times 10^{-7}\; (4.1\,\times 10^{-7}).$
\end{center}
We also used a Bayesian approach with flat prior and Gaussian (smeared) uncertainties~\cite{bayes} and
obtained the limit of ${\cal B}(B^0_s~\rightarrow~\mu^+~\mu^-) \leq 5.1\, \times 10^{-7}\; (4.1\,\times 10^{-7})$
at the 95\% (90\%) C.L. This new result is presently the most stringent bound on ${\cal B}(B^0_s \rightarrow \mu^+ \mu^-)$, improving the previously published
value~\cite{cdfII} and can be used to constrain models of new physics beyond the SM.

%
We thank the staffs at Fermilab and collaborating institutions, 
and acknowledge support from the 
Department of Energy and National Science Foundation (USA),  
Commissariat  \` a l'Energie Atomique and 
CNRS/Institut National de Physique Nucl\'eaire et 
de Physique des Particules (France), 
Ministry of Education and Science, Agency for Atomic 
   Energy and RF President Grants Program (Russia),
CAPES, CNPq, FAPERJ, FAPESP and FUNDUNESP (Brazil),
Departments of Atomic Energy and Science and Technology (India),
Colciencias (Colombia),
CONACyT (Mexico),
KRF (Korea),
CONICET and UBACyT (Argentina),
The Foundation for Fundamental Research on Matter (The Netherlands),
PPARC (United Kingdom),
Ministry of Education (Czech Republic),
Natural Sciences and Engineering Research Council and 
WestGrid Project (Canada),
BMBF and DFG (Germany),
A.P.~Sloan Foundation,
Research Corporation,
Texas Advanced Research Program,
and the Alexander von Humboldt Foundation.
%


\begin{thebibliography}{99}
\bibitem{conjugated} Charge conjugated states are included implicitly.
\bibitem{eff_ratio}The ratio
  $\epsilon_{\mu\mu}^{B^0_d}/\epsilon_{\mu\mu}^{B^0_s}$ has been
  determined from simulation to be $0.92\pm 0.04$, with the uncertainty due to limited MC statistics.
\bibitem{norm} In addition to $B^{\pm}\rightarrow J/\psi\,K^{\pm}$,
  the other possible normalization channel is $B_s^0\rightarrow
  J/\psi\,\phi$. We have not used it due to low statistics, a large uncertainty on its branching fraction and a poorly known mixture of CP even and CP odd decay modes with lifetime differences.
\bibitem{limit} This limit is derived with the world average fragmentation value~\cite{pdg} of $f_{b\rightarrow B^0_{s}}/f_{b \rightarrow B_{u,d}}=0.270\pm 0.034$. A fragmentation ratio based on Tevatron data alone~\cite{cdf_frag}, but with larger uncertainty, gives $f_{b\rightarrow B^0_{s}}/f_{b \rightarrow B_{u,d}}=0.42\pm 0.14$ and results in a limit of $4.1\times 10^{-7}$ ($3.1\times 10^{-7}$) at a 95\% (90\%) C.L. using the method of Ref.~\cite{conrad}.
\end{thebibliography}
\end{document}